\documentclass{mem}
\usepackage{natbib}
\usepackage{txfonts}
\usepackage{balance}
\usepackage{graphicx}
\usepackage{psfig}
\usepackage[a4paper]{hyperref}
\idline{1}{1}

\begin{document}
\def\teff{$T\rm_{eff }$}
\def\kms{$\mathrm {km s}^{-1}$}
\input{psfig}

\title{Extragalactic jets: a new perspective}
   \subtitle{}

\author{
Gabriele Ghisellini 
          }

  \offprints{G. Ghisellini}

\institute{
Istituto Nazionale di Astrofisica --
Osservatorio Astronomico di Brera  \\
Via Bianchi 46, I--23807 Merate, Italy
\email{gabriele.ghisellini@brera.inaf.it}
}

\authorrunning{G. Ghisellini}

\titlerunning{Extragalactic jets: a new perspective}

\abstract{
The power carried by the jet of blazars is large, compared to the
luminosity produced by their accretion disk, and is probably
in the form of kinetic energy of a normal electron--proton plasma.
The Poynting flux is modest,
as suggested by the inconspicuous synchrotron luminosity
when compared to the high energy (hard X--rays and $\gamma$--ray) one,
assumed to be produced by the inverse Compton process.
%
It is suggested that the jet power and the SED 
(Spectral Energy Distribution) of its emission 
are linked to the mass of the black hole $M$ and the accretion rate $\dot M$.
This corresponds to a new ``blazar sequence" based on $M$ and $\dot M$
instead of only the observed blazar luminosity.
These ideas can be tested quite easily once the AGILE and especially 
the GLAST satellite observations, coupled with information in 
the optical/X--ray band from Swift, will allow the knowledge 
of the entire SED  of hundreds blazars.
}

\maketitle{}

\section{Introduction}

The power and the content of relativistic extragalactic jets is a long 
standing issue.
The power can only be measured indirectly: through minimum energy 
arguments coupled with estimates of the lifetime of radio--lobes, thought to act
as calorimeters (i.e. \cite{Burbidge58}, \cite{RS91}); or through the
work done by X--ray cavities corresponding to radio ``bubbles"
(e.g. \cite{Allen06}); or using the observed luminosities to
infer the bulk motion and the physical properties of the emitting
plasma, from the scale of hundreds of kpc 
(``Chandra'' jets,  e.g. \cite{CGT01}, \cite{T00}, \cite{GC01}), to the
VLBI scale (\cite{CF93}),
and to the $\gamma$--ray zone, at the sub--pc scale (\cite{CG08},
hereafter CG08).
The latter method, besides the size of the source
and the bulk Lorentz factor, allows the determination of the magnetic field
strength and the number of the emitting electrons.
It is then possible to infer, separately, the power associated with the
kinetic motion of the relativistic electrons (or pairs), the Poynting flux,
and the power associated with the emitted radiation.
Then we can get a clue on the content of the jet, namely if we need protons
carrying most of the kinetic energy or if electrons and/or magnetic fields
are enough.

Armed with these new information on the jet power, we can 
explore other fundamental issues, such as the jet/accretion 
disk connection.
We will find that the jet is more powerful than the 
radiation emitted by the accretion disk, by approximately
one order of magnitude, in agreement with earlier findings
(e.g. CG08; \cite{MT03}).
There is an approximately linear relation between them, at least for 
Flat Spectrum Radio Quasars (FSRQs), where the accretion
luminosity, and/or the broad lines, are visible.
This hints to a direct link between the jet power $P_{\rm jet}$ and
the accretion rate $\dot M$.

Another important piece of information concerns the location where
most of the jet radiation is produced.
Contrary to earlier ``continuous'' models (e.g. \cite{M80}) we now know that
the jet dissipates in a well localized region, not too close to the 
black hole and accretion disk (to avoid having the $\gamma$--rays absorbed in
$\gamma$--$\gamma\to e^\pm$ collisions; \cite{GM96}), 
and not too far, to account for the fast observed variability:
a dissipation region at a distance $R_{\rm diss}$ of
a few hundreds of Schwarzshild radii is appropriate.
Thus $R_{\rm diss}$ should be related to the black hole mass $M$,
that can now be estimated through a number of empirical relations.

\begin{figure}[t!]
\vskip -0.7 true cm
\hskip -0.5 true cm
\psfig{figure=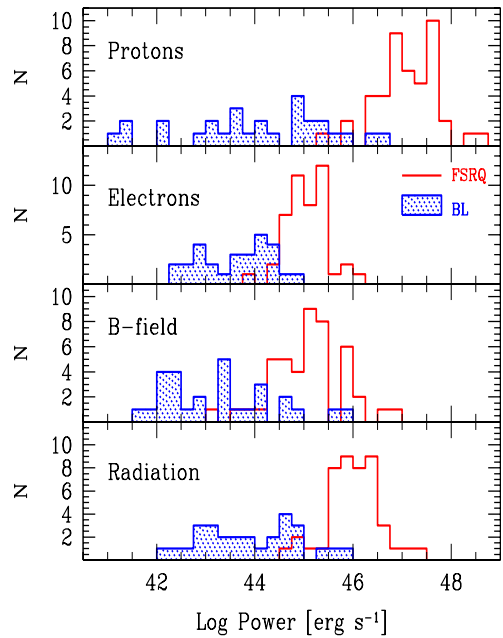,width=7.5cm,height=10cm}
\vskip -0.7 true cm
\caption{
\footnotesize
Power associated to protons, emitting electrons (or
e$^{\pm}$), Poynting flux and radiation for a sample of
blazar (both FSRQs and BL Lacs).
Hatched areas correspond to BL Lacs.  
The power in protons is derived assuming one (cold) proton per emitting electron.
Adapted from \cite{CG08}.
}
\label{power}
\end{figure}

Finally, the size $R_{\rm BLR}$ of the Broad Line Region (BLR) is related to
the disk luminosity $L_{\rm disk}$.
Although this is a still debated issue (\cite{Bentz06}; \cite{Kaspi07}) 
we approximately have $R_{\rm BLR}\propto L_{\rm disk}^{1/2}$,
indicating a constant ionization parameter, as expected.

These findings allows us to construct a new sequence
for blazars (\cite{GT08}, hereafter GT08), 
based on $M$ and $\dot M$, that extends
and completes our earlier ``spectral blazar sequence''
(\cite{Fossati98}; \cite{Ghisellini98}) that was based on 
the observed bolometric luminosity.

\section{Jet power}

Fig. \ref{power} shows the power of blazar jets as found by CG08,
in the form of cold protons, emitting leptons, Poynting flux and radiation.
The hatched areas corresponds to BL Lac objects.
Consider first FSRQs. 
The power in radiation, $P_{\rm rad}$, is directly related to the observed 
(bolometric) luminosity $L^{\rm obs}_{\rm rad}$ by 
$P_{\rm rad} \sim L^{\rm obs}_{\rm rad}/\Gamma^2$.
Apart from the bulk Lorentz factor $\Gamma$, $P_{\rm rad}$ 
is model--independent, thus a rather robust estimate.
It is larger than the power in emitting 
electrons or in Poynting flux.
This shows that: 
\begin{enumerate}
\item the jet cannot be made only by emitting $e^\pm$
pairs, since their kinetic power is less then $P_{\rm rad}$; 
\item the magnetic field cannot be dynamically important, at least
where most of the flux is produced, since the Poynting flux is less than $P_{\rm rad}$; 
\item the jet must carry most of its power in some other form. 
\end{enumerate}
The simplest possibility is that the jet, besides leptons,
carry also (cold) protons: one per emitting electron solves the power budget,
and leaves enough power to be carried to the radio lobes.
 
Note that BL Lac objects are on average much less powerful than FSRQs, 
but similar arguments holds also for them (see the discussion in CG08).

\begin{figure}[t!]
\vskip -0.5 true cm
\hskip -0.3 true cm
\psfig{figure=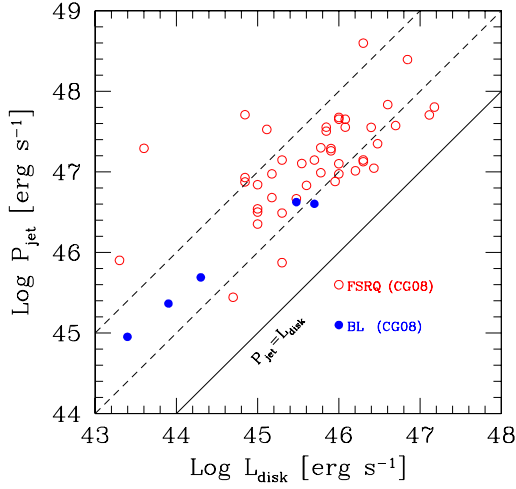,width=7.3cm,height=7.3cm}
\vskip -0.3 true cm
\caption{
\footnotesize
The jet power of the blazar studied in CG08 as a function
of their accretion disk luminosity (as derived the fitting
their SED).
The (few) filled circles correspond to BL Lac objects,
the empty circles to FSRQs.
}
\label{ldlj}
\end{figure}

Fig. \ref{ldlj} shows the jet power (including protons)
as a function of the accretion disk luminosity.
On average, the jet is one order of magnitude more powerful
than $L_{\rm disk}$.
It is fair to say that the jet power is derived 
by modelling their SED, and especially the high energy $\gamma$--ray
emission (indeed, CG08 studied a $\gamma$--ray selected sample of blazars).
There is then a bias favoring blazars in high $\gamma$--ray states,
hence the derived jet power corresponds to states brighter than average.

\section{The blazar spectral sequence}

\begin{figure}[t!]
\vskip -0.5 true cm
\hskip -0.3 true cm
\psfig{figure=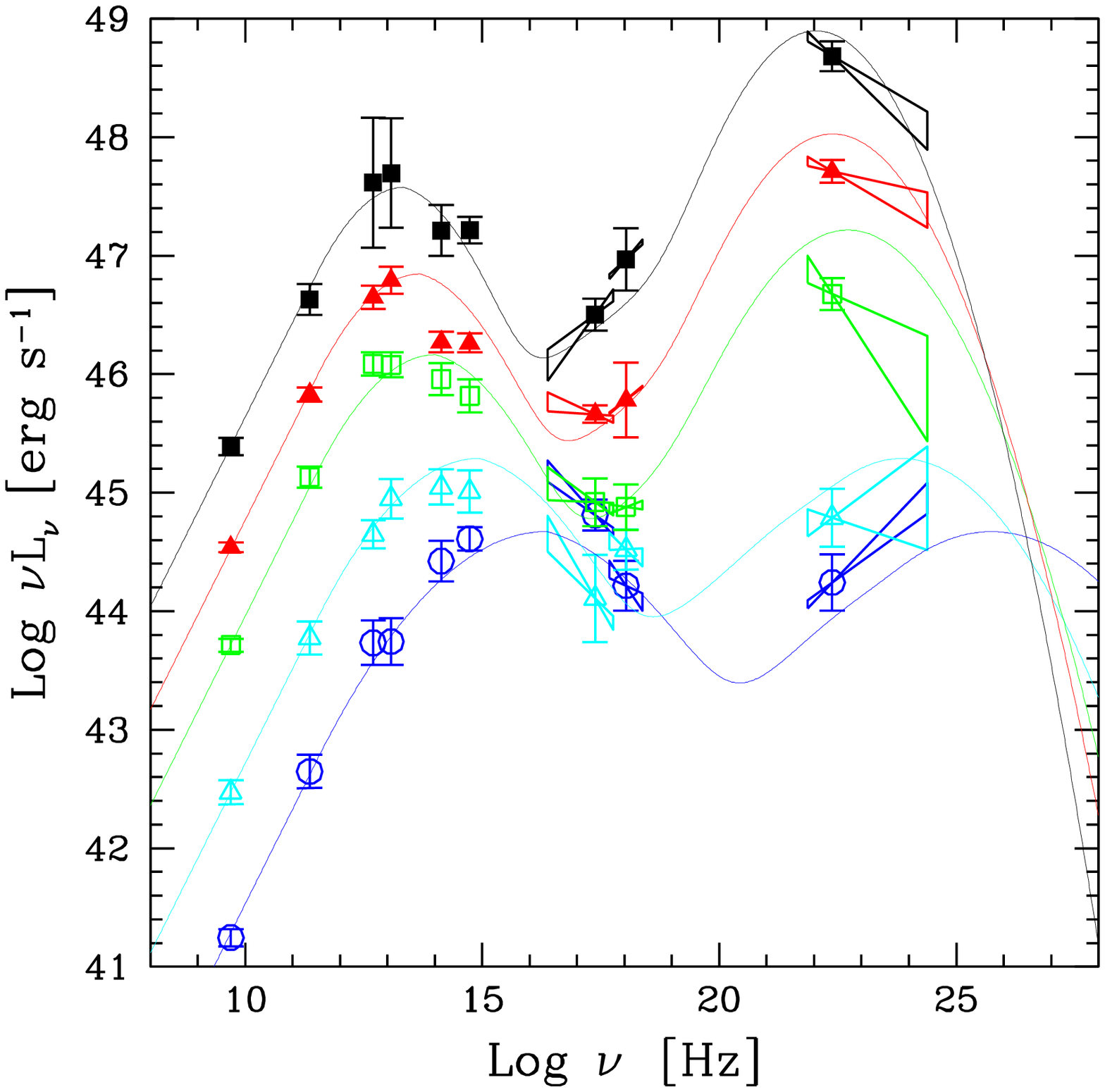,width=7.3cm,height=7.3cm}
\vskip -0.5 true cm
\caption{
\footnotesize
The spectral blazar sequence 
(adapted from \cite{Fossati98} and \cite{Donato01}).
}
\label{sequence}
\end{figure}

The SED of blazars if characterized by two broad peaks
(in the mm--soft X--ray and in MeV--TeV energy ranges),
thought to be produced by synchrotron and inverse Compton
emission by the same electrons.
\cite{Fossati98} noted that blazars form a sequence,
shown in Fig. \ref{sequence}:
as the luminosity increases, the peak frequencies of the two
broad humps shift to smaller values, and the high energy
peak becomes more dominant.
This was interpreted (\cite{Ghisellini98}) as due to different
amount of radiative cooling suffered by the emitting electrons.
In powerful sources (FSRQs) the cooling is more severe, implying
typical electron energies smaller than in less powerful sources (BL Lacs).
Furthermore, the presence of broad lines in powerful
FSRQs (and the absence in BL Lacs) makes the
inverse Compton process in FSRQs more important than in BL Lacs.
This phenomenological 
blazar sequence is controlled by one parameter:
the bolometric observed luminosity.

\section{A new blazar sequence}

As mentioned in the introduction, GT08 
proposed a new sequence of blazars,
based on two parameters: $M$ and $\dot M$.
The aim was not only to link the general properties
of blazar jets (i.e. their power) to accretion, but
more specifically to construct a scenario in which 
the SED of a blazar is predictable knowing $M$ and $\dot M$.
Vice--versa, knowing in detail the SED, we can estimate the
black hole mass and the accretion rate.
This (ambitious) program is possible only if some drastic
simplification is made, resulting in an {\it average} description
of the blazar itself.

\subsection{Assumptions}

This key assumptions are:
\begin{itemize}

\item The total jet power $P_{\rm jet}$ is always proportional to the
accretion rate: $P_{\rm jet} =\eta_{\rm j} \dot M c^2$.
The efficiency factor is $\eta_{\rm j} \sim 0.3-0.5$, larger than the corresponding
accretion efficiency for a standard disk.

\item Relativistic electrons and the Poynting flux carry, respectively, a fraction
$\epsilon_{\rm e}$ and $\epsilon_{\rm B}$ of the total jet power $P_{\rm jet}$.
$\epsilon_{\rm e}$ and $\epsilon_{\rm B}$ are power law functions of 
$P_{\rm jet}$, derived empirically through existing observations and 
analysis (see CG08).

\item The dissipation radius is proportional to $M$ (i.e. $R_{\rm diss}$ is
always of the order of 100--300 Schwarzschild radii).
Since we assume a conical jet with semiaperture angle $\psi_{\rm j}=0.1$,
the size of the emitting region is $R=\psi_{\rm j}R_{\rm diss}=R_{\rm diss}/10$.

\item The BLR has a size $R_{\rm BLR} \propto L_{\rm disk}^{1/2}$,
but it does not exist below a critical value of $L_{\rm disk}/L_{\rm Edd}$
($L_{\rm Edd}$ is the Eddington luminosity).
Below this critical value 
the accretion changes regime, becoming radiatively inefficient.
Sources with disks emitting below this value are BL Lac objects 
(and FR I radio--galaxies, see \cite{GC01}).

\item We can find the shape and normalization of the 
emitting electron distribution taking into account their
radiative cooling and following some simple prescriptions
(as discussed in GT08).

\item As for the radiation processes, we assume synchrotron and inverse Compton.
For the Compton scattering, we use synchrotron as well broad line
photons, but we neglect the latter when $R_{\rm diss}>R_{\rm BLR}$.

\end{itemize}
\begin{figure}[t!]
\vskip -0.7 true cm
\hskip -1.8 true cm
\psfig{figure=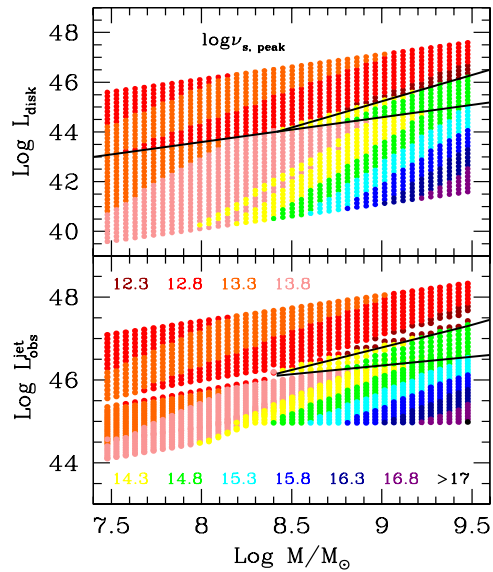,width=9.5cm,height=9cm}
\vskip -0.5 true cm
\caption{
\footnotesize 
In the luminosity--black hole mass plane blazars with different
synchrotron peak frequencies occupy different regions,
as illustrated by the different levels of grey.
The labels show the corresponding values of $\log\nu_{\rm S}$.
In the top panel we show the distribution of $\nu_{\rm S}$ values
in the $L_{\rm disk}$--$M$ plane.
Bottom: the same for the $L^{\rm obs}_{\rm jet}$--$M$ plane. 
}
\label{vs}
\end{figure}
\begin{figure}[t!]
\vskip -0.7 true cm
\hskip -1.8 true cm
\psfig{figure=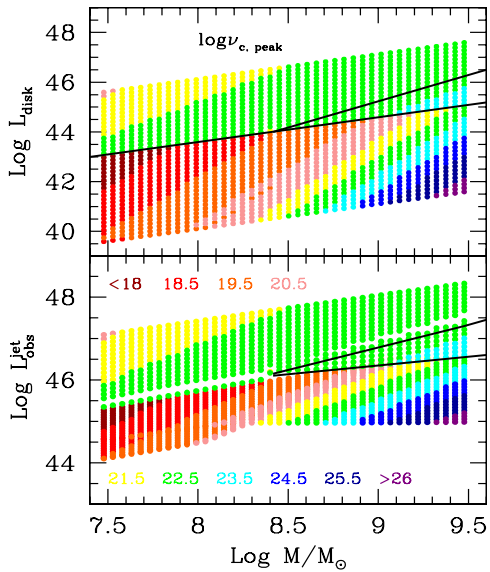,width=9.5cm,height=9cm}
\vskip -0.5 true cm
\caption{
\footnotesize
Same as in Fig. \ref{vs}, but for the Compton peak frequency $\nu_{\rm C}$.
}
\label{vc}
\end{figure}

\subsection{Results}

We call ``blue'' (``red'') a blazar with relatively large (small)
synchrotron ($\nu_{\rm S}$) and inverse Compton ($\nu_{\rm C}$) peak frequencies.
Just to fix the ideas, blazars with $\nu_{\rm S}>10^{15}$ Hz and 
$\nu_{\rm C}>10^{24}$ Hz are blue.
There are a few important and immediate consequences born out from our scheme.

The dissipation region is proportional to $M$, the size of the BLR
depends on $L_{\rm disk}$.
Sources with large $M$ and small $L_{\rm disk}$ (but above the critical value),
can then have $R_{\rm diss}$ beyond the BLR.
These jets are also not particularly powerful, since $\dot M$ is
relatively small.
In these conditions the radiative cooling of the emitting electrons
is modest, both because the magnetic field is small and because 
the seed photons from the BLR are negligible as seed photons for 
the Compton process.
This implies that the energy of those electrons emitting at the 
peaks of the SED are very large: we have a blue blazar.
But since we do observe broad lines from this source, we will classify it
as a FSRQ.
Then {\it blue quasars may exist} (and ``live'' beetween the solid
black lines of Fig. \ref{vs} and Fig. \ref{vc}).

Another simple consequence of our scheme concerns sources with small $M$,
presumably the most numerous.
They will have, on average, a smaller disk luminosity and jet power
then the high--$M$ counterparts.
Being fainter, we could confuse them with (slightly) misaligned blazars,
although the latter should be characterized by a more prominent blue bump.
For $\dot M$ above the critical value, the radiative cooling is
severe, and the blazar would be red.
Then {\it red and relatively low power blazars should exist}.

This conclusions stem out directly from our key assumptions,
but to go further we need to derive the {\it Compton dominance}
of blazars, that is the ratio between the inverse Compton
and the synchrotron luminosities.

\begin{figure}[t!]
\vskip -0.6 true cm
\hskip -1.8 true cm
\psfig{figure=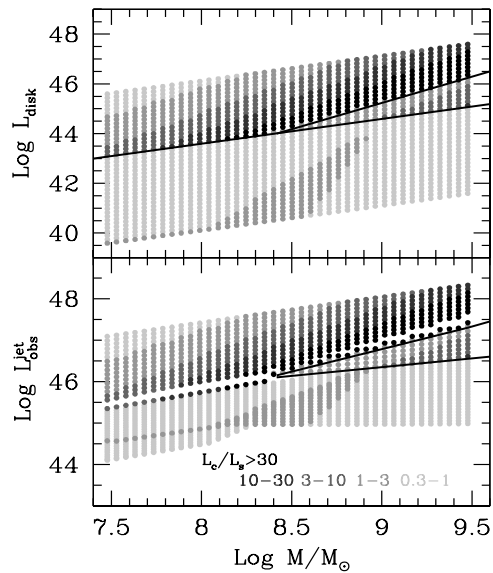,width=9.5cm,height=9cm}
\vskip -0.3 true cm
\caption{
\footnotesize
The disk accretion luminosity (top panel) and the observed 
bolometric jet emission (bottom panel), as a function of black hole mass.
The solid lines on both panels correspond to 
$L/L_{\rm Edd}=3\times 10^{-3}$ (below this line we have BL Lacs, above we have FSRQs) 
and to $R_{\rm diss}>R_{\rm BLR}$ (corresponding to the region of blue FSRQs).
The different grey levels correspond to different ``Compton dominances",
i.e. different Compton to synchrotron luminosity ratios $L_C/L_S$. From \cite{GT08}.
}
\label{ml}
\end{figure}

Consider first FSRQs, i.e. blazar with broad emission lines.
Fig. \ref{ml} shows, with different levels of grey, the Compton
dominance for objects with given $M$ and $L_{\rm disk}$ 
of $L_{\rm jet}^{\rm obs}$.
The diagonal solid line corresponds to $L_{\rm disk}/L_{\rm Edd}$ greater
than a critical value, below which the accretion becomes radiatively
inefficient.
FSRQs lie above this line; BL Lacs lie below.
As a rule, when $R_{\rm diss} <R_{\rm BLR}$,
the smaller the magnetic field $B$, the larger the Compton 
dominance, since the BLR photons dominate the cooling
through the Compton process. 
We have small values of $B$ for larger $R_{\rm diss}$
(i.e. larger black hole masses) and small jet powers.
Vice--versa, for small $M$ and large jet powers, we have 
large $B$--values: these are very red blazars (Fig. \ref{vs}), 
but not extreme in Compton dominance (Fig. \ref{ml}).

Consider now BL Lac objects: there is no BLR for them, and
the Compton dominance is related to $\epsilon_{\rm e}/\epsilon_{\rm B}$,
as in Gamma Ray Bursts.
As a rule, the lack of broad line photons makes these sources
less Compton dominated than FSRQs, as it is clearly visible in Fig. \ref{ml}.
In turn, this makes BL Lacs bluer than FSRQs, since the lack
of BLR photons implies less cooling.

\section{Implications for cosmic evolution}

Our scheme bears important implications for the evolutionary 
properties of blazars, if the accretion rate in jetted sources 
evolves as in radio--quiet AGNs. 
Assume, for simplicity, that the black hole mass function 
does not evolve with redshift.
If larger $\dot M$ were more common in the past, 
a larger fraction of sources were above the critical accretion rate
to have the BLR. 
So FSRQs (and their parent FR II radio--galaxies), 
should evolve {\it positively} with redshift.
Vice--versa, BL Lacs (and FR I) should be more common now, and rarer in the past,
i.e. they should show a {\it negative} evolution.
This is in agreement to what proposed by \cite{Cavaliere02}.
Taken altogether, blazars should then show a similar evolution than radio--quiet
quasars (see also \cite{Maraschi94}), 
but with the caveat that they probably have larger black hole masses.

\section{Conclusions}

When a blazar is flaring or it is bright in $\gamma$--rays, its 
jet carries a power that is a factor $\sim$10 greater than the luminosity produced 
by the accretion disk (and even more in BL Lacs).
Taking a duty cycle (fraction of time spent in high state)
of order $\sim$0.1, we obtain $\langle P_{\rm jet}\rangle \approx L_{\rm disk}$.
In FSRQs, this power is mainly in the form of bulk motion of protons;
relativistic electrons, $e^\pm$ pairs and magnetic fields are not dynamically important,
confirming earlier findings by \cite{Sikora00}.
Since, in FSRQs whose disk is an efficient radiator, 
$P_{\rm jet}\propto L_{\rm disk}$, there must be a strong link between 
jet power and accretion rate.
This motivated us to assume, {\it for all blazars}
$P_{\rm jet} \propto \dot M c^2$.
This is the ansatz at the heart of the new blazar sequence we are suggesting,
relating the blazar SED to $M$ and $\dot M$.
GLAST, with the help of the AGILE and the Swift satellites, will hopefully provide
a good spectral coverage and really simultaneous SED for hundreds of blazars.
This, together with the knowledge of the black hole mass and
accretion rate (through emission lines or through direct observations
of the blue bump) makes our scenario easily testable.

\begin{acknowledgements}
I thank A. Celotti, L. Maraschi and F. Tavecchio for discussions. 
\end{acknowledgements}

\bibliographystyle{aa}

\end{document}